\documentclass[10pt,iop]{emulateapj}

\usepackage{graphicx}
\usepackage{float}
\usepackage{amsmath}
\usepackage{epsfig}

\renewcommand{\P}{\widetilde{P}}

\newcommand{\ve}[1]{\mathbf{#1}}
\renewcommand{\L}{\ve{\Lambda}}

\newcommand{\A}{\ve{A}}
\newcommand{\Y}{\ve{Y}}
\newcommand{\N}{\ve{N}}

\renewcommand{\S}{\ve{S}}

\newcommand{\s}{\ve{s}}

\newcommand{\n}{\ve{n}}
\renewcommand{\d}{\ve{d}}

\newcommand{\lm}{{\ell m}}
\newcommand{\lmp}{{\ell' m'}}

\newcommand{\Npix}{{N_{\text{pix}}}}
\newcommand{\Nside}{N_{\text{side}}}

\newcommand{\mat}[1]{\left[ \begin{array}{ccccccccccccccc} #1 \end{array} \right]}

\newcommand{\lmax}{L}

\newcommand{\cmt}[1]{}

\begin{document}

\title{Wavemoth -- Fast spherical harmonic transforms by butterfly
  matrix compression}

\author{D. S. Seljebotn\altaffilmark{1}}

\email{d.s.seljebotn@astro.uio.no}

\altaffiltext{1}{Institute of Theoretical Astrophysics, University of
  Oslo, P.O.\ Box 1029 Blindern, N-0315 Oslo, Norway}

\begin{abstract}
  We present {\em Wavemoth}, an experimental open source code for
  computing scalar spherical harmonic transforms (SHTs).  Such
  transforms are ubiquitous in astronomical data analysis.  Our code
  performs substantially better than existing publicly available codes
  due to improvements on two fronts. First, the computational core is
  made more efficient by using small amounts of precomputed data, as
  well as paying attention to CPU instruction pipelining and cache
  usage.  Second, Wavemoth makes use of a fast and numerically stable
  algorithm based on compressing a set of linear operators in a
  precomputation step. The resulting SHT scales as $O(\lmax^2 \log^2
  \lmax)$ for the resolution range of practical interest, where
  $\lmax$ denotes the spherical harmonic truncation degree. For low
  and medium-range resolutions, Wavemoth tends to be twice as fast as
  {\em libpsht}, which is the current state of the art implementation
  for the HEALPix grid. At the resolution of the Planck experiment,
  $\lmax \sim 4000$, Wavemoth is between three and six times faster
  than libpsht, depending on the computer architecture and the
  required precision.  Due to the experimental nature of the project,
  only spherical harmonic synthesis is currently supported, although
  adding support for spherical harmonic analysis should be trivial.

\end{abstract}
\shorttitle{Wavemoth: Fast spherical harmonic transforms}
\section{Background}
\label{sec:introduction}
The spherical harmonic transform (SHT) is the spherical analog of the
Fourier transform, and is an essential tool for data analysis and
simulation on the sphere. A scalar field $f(\theta, \phi)$ on the unit
sphere can be expressed as a weighted sum of the spherical harmonic
basis functions $Y_\lm(\theta, \phi)$,
\begin{equation}
  \label{eq:shdef}
    f(\theta, \phi) = \sum_{\ell = 0}^\infty \sum_{m=-\ell}^\ell a_\lm
      Y_\lm(\theta, \phi).
\end{equation}
The coefficients $a_\lm$ contain the spectral information of the
field, with higher $\ell$ corresponding to higher frequencies. In
calculations the spherical harmonic expansion is truncated for $\ell >
\lmax$, and the spherical field represented by $O(\lmax^2)$ grid
samples. Computing the sum above is known as the {\em backward SHT} or
{\em synthesis}, while the inverse problem of finding the spherical
harmonic coefficients $a_\lm$ given the field $f$ is known as the {\em
  forward SHT} or {\em analysis}.

In order to compute an SHT, the first step is nearly always to employ
a separation of sums, which we review in Section \ref{sec:FFT}, to
decrease the cost from $O(\lmax^4)$ to $O(\lmax^3)$. We will refer to
codes that take no measures beyond this to reduce complexity as
brute-force codes.  Of these, HEALPix \citep{healpix} is one very
widely used package, in particular among CMB researchers.

Recently, the libpsht package \citep{libpsht} halved the computation
time with respect to the original HEALPix implementation, simply
through code optimizations. As of version 2.20, HEALPix uses libpsht
as the backend for SHTs.  Other packages using the brute-force
algorithm include S$^2$HAT \citep{s2hat-a,s2hat-b}, focusing on
cluster parallelization and implementations on the GPU, as well as GLESP
\citep{doi:my/GLESP} and ssht \citep{doi:10.1109/TSP.2011.2166394},
focusing on spherical grids with more accurate spherical harmonic
analysis than what can be achieved on the HEALPix grid.

The discovery of Fast Fourier Transforms (FFTs) has been all-important
for signal analysis over the past half century, and there is no lack
of high quality commercial and open source libraries to perform FFTs
with stunning speed.  Unfortunately, the straightforward
divide-and-conquer FFT algorithms do not generalize to SHTs, and
research in fast SHT algorithms has yet to reach maturity in the sense
of widely adopted algorithms and libraries.

The libftsh library \citep{mohlenkamp} uses local trigonometric
expansions to compress the spherical harmonic linear operator,
resulting in a computational scaling of $O(\lmax^{5/2}\log \lmax)$ in
finite precision arithmetic.  SpharmonicKit
\citep{doi:10.1007/s00041-003-0018-9} implements a divide-and-conquer
scheme which scales as $O(\lmax^2 \log^2 \lmax)$.  We comment further
on these in Section \ref{sec:comparison-with-others}.  Other
algorithms have also been presented but either suffer from problems
with numerical stability, are impractical for current resolutions, or
simply lack publicly available implementations
\citep[e.g.,][]{doi:my/suda,doi:10.1016/S0377-04270300546-6,tygert1,tygert2,tygert3}.

We present Wavemoth\footnote{http://github.com/wavemoth; commit
  59ec31b8 was used to produce the results of this paper.}, an
experimental open source implementation of the algorithm of
\cite{tygert3}.  This algorithm has several appealing features.
First, it is simple to implement and optimize. Second, it is
inherently numerically stable. Third, its constant prefactor is
reasonable, yielding substantial gains already at $\lmax \sim 2000$.
The accuracy of the algorithm is finite, but can be arbitrarily
chosen. For any given accuracy, the computational scaling is
$O(\lmax^2 \log^2 \lmax)$, but lowering the requested accuracy makes
the constant prefactor smaller.

We stress that our work consists solely in providing an optimized
implementation. While we review the basics of the algorithm in Section
\ref{sec:fast-legendre}, \cite{tygert3} should be consulted for
details and proofs.  We have focused in particular on the HEALPix
grid, and use libpsht as our baseline for comparisons. However, all
methods work equally well for any other grid with iso-latitude rings.

Section \ref{sec:baseline} reviews SHTs in more detail, as well as the
computational methods that are widely known and used across all
popular codes. Section \ref{sec:fast-legendre} reviews the algorithm
of \cite{tygert3} and how we have adapted it to our purposes. Section
\ref{sec:implementation} focuses on the high-level aspects of software
development and provides benchmarks, while an appendix provides the
low-level implementation details.

\section{Baseline algorithms}
\label{sec:baseline}

\subsection{The spherical harmonic basis functions}
\label{sec:backward}
We use the convention that points on the sphere are parameterized by a
co-latitude $\theta \in [0, \pi]$, where $0$ corresponds to the
``north pole'', and a longitude $\phi \in [0, 2\pi)$.  The spherical
harmonic basis functions $Y_\lm(\theta, \phi)$ can then be expressed in
terms of the {\em associated Legendre functions} $P_\ell^m(z)$. Assuming $m \ge 0$, we have
\begin{equation}
  \label{eq:Y-def}
  \begin{split}
      Y_\lm(\theta, \phi) &=
\sqrt{\frac{2\ell + 1}{4\pi} \frac{(\ell - m)!}{(\ell + m)!}}
 P^{m}_\ell (\cos \theta)
e^{i m \phi} \\
    &\equiv \P^{m}_\ell (\cos \theta)  e^{i m \phi},
  \end{split}
\end{equation}
where we define the {\em normalized associated Legendre function}
$\P^{m}_\ell$. Our definition follows that of \cite{NR}; the
normalization differs by a factor of $\sqrt{1/2}$ from the one in
\cite{tygert3}.

Note that while the spherical harmonics $Y_\lm$ and the coefficients
$a_\lm$ are complex, $\P^{m}_\ell$ is real for the argument range of
interest. For negative $m$, the symmetry $Y_{\ell,-m} = (-1)^mY_\lm^*$
can be used, although this is only needed for complex fields. Wavemoth
only supports real fields, which have spherical harmonic expansions
obeying $a_\lm = (-1)^m a_{\ell-m}^*$.

\subsection{Discretization and the forward transform}
\label{sec:analysis}
For computational work one has to assume that one is working with a
band-limited signal, so that $a_\lm = 0$ when $\ell > \lmax$.
The SHT synthesis is then given simply by evaluating
equation \eqref{eq:shdef} in a set of points on the sphere.

The opposite problem of computing $a_\lm$ given $f(\theta_j, \phi_j)$,
namely spherical harmonic analysis, is less straightforward. In the
limit of infinite resolution, we have
\begin{equation}
  \label{eq:sh-forward}
  a_\lm = \int f(\theta, \phi) Y_\lm^*(\theta, \phi) d\Omega,
\end{equation}
where $d\Omega$ indicates integration over the sphere.
This follows easily from the orthogonality property,
\begin{equation}
\label{eq:orthogonality}
\int Y_\lm Y_\lmp^* d\Omega = \delta_{\ell \ell'} \delta_{m m'}.
\end{equation}
There is no canonical way of choosing sample points on the sphere.
The simplest grid conceptually is the equiangular grid.
\cite{doi:my/GLESP} and \cite{doi:10.1109/TSP.2011.2166394} describe
grids that carries the orthogonality property of the
continuous spherical harmonics over to the discretized operator. In
contrast, the HEALPix grid \citep{healpix} trades orthogonality for
the property that each pixel has the same area, which is convenient
for many operations in the pixel basis.

Independent of what grid is chosen, a natural approach to spherical
harmonic analysis is to use a quadrature rule with some weights $w_j$,
so that
\begin{equation}
  \label{eq:sh-forward-quadrature}
  a_\lm = \sum_{j=1}^\Npix w_j f(\theta_j, \phi_j) Y_\lm^*(\theta_j, \phi_j).
\end{equation}
On the HEALPix grid the numerical accuracy of this approach is
limited, but it is still the most common procedure.

Some real world signal analysis problems do not need the forward
transform at all.  In the presence of measurement noise in the pixel
basis, one can argue that the best approach is not to pull the noise
part of the signal into spherical harmonic basis at all.
For instance, consider the archetypical CMB data model,
\begin{equation}
  \label{eq:datamodel}
  \d = \Y\s + \n,
\end{equation}
where $\d$ represents a vector of pixels on the sky with observed data
(not necessarily the full sky), $\s$ represents our signal of interest
in spherical harmonic basis, and $\n$ represents instrumental noise in
each pixel.  Spherical harmonic synthesis is denoted $\Y$; note
that equation \eqref{eq:shdef} describes a linear operator and can be
written $\ve{f} = \Y \ve{a}$.

If we now assume that $\s$ and $\n$ are Gaussian random vectors with
vanishing mean and known covariance matrices $\S$ and $\N$,
respectively, then the maximum likelihood estimate of the signal
is given by
\begin{equation}
  \label{eq:wiener}
  \hat{\s} = (\S^{-1} + \Y^\dagger \N^{-1} \Y)^{-1} \Y^\dagger \N^{-1} \d,
\end{equation}
with $\hat{\s}$ in spherical harmonic basis. This system can be solved
with reasonable efficiency by iterative methods. Note that we are here
only concerned with the effect of $\Y$ as a non-invertible projection,
and that no spherical harmonic analysis is ever performed, only the
adjoint synthesis. Thus, neither the non-orthogonality caused by the
HEALPix grid, nor masking out large parts of the sky, is a concern.
See \cite{doi:10.1086/525277} and references therein for more details
on this technique in the context of CMB analysis.

\subsection{Applying the Fast Fourier Transform}
\label{sec:FFT}
The first step in speeding up the spherical harmonic transform beyond
the $O(\lmax^4)$ brute-force sum is a simple
separation of sums. For this to work well, pixels must be arranged on
a set of iso-latitude rings, with equidistant pixels within each
ring. All grids in use for high-resolution data has this property.

We show the case for SHT synthesis; analysis can be treated in
the same way. Starting from equation \eqref{eq:shdef}, we have, for
pixel $j$ within ring $k$, and with $z_k \equiv \cos \theta_k$,
\begin{equation}
  \label{eq:sh-baseline}
  \begin{split}
    f(\theta_{k},\phi_{k,j})
      &= \sum_{m=-\lmax}^\lmax \left[ 
          \sum_{\ell = |m|}^\lmax a_\lm \P^{m}_\ell(z_k)
          \right] e^{i m \phi_{k,j}} \\
      &\equiv \sum_{m=-\lmax}^\lmax q_m(z_k) e^{i m \phi_{k,j}},
    \end{split}
\end{equation}
where we introduce $q_m(z_k)$. Assuming that ring $k$ contains $J_k$
pixels, their equidistant longitude is given by
\begin{equation}
  \label{eq:phi}
  \phi_{k,j} = \phi_{k,0} + \frac{2 \pi j}{J_k}.
\end{equation}
Since $e^{ix}$ has period $2\pi$, and since $q_m(z_k) = 0$ whenever $|m| > \lmax$, we find that
\begin{equation}
  \label{eq:pad-and-wrap}
  \begin{split}
\sum_{m=-\lmax}^\lmax q_{k,m} e^{i m \phi_{k,j}} = 
\sum_{j=0}^{J_k - 1} \tau_{j}(z_k) e^{2\pi j i / J_k}
    \end{split}
\end{equation}
with
\begin{equation}
  \label{eq:tau}
\tau_{j}(z_k) = \sum_{t=-\infty}^\infty q_{J_k t + j}(z_k)  e^{i\phi_{k,0}(J_k t + j)}.
\end{equation}
Thus one can phase-shift the coefficients $q_{m}(z_k)$ to match the
ring grid, wrap around or pad with zeros, and perform a regular
backward FFT.  The symmetries of the spherical harmonic
coefficients of a real field carry over directly to the Hermitian
property of real Fourier transforms.

This separation of sums represents a first step in speeding up the
SHT, and is implemented in all packages for high-resolution spherical
harmonic transforms.

\subsection{Legendre transforms and even/odd symmetry}
\label{sec:even-odd-symmetry}
\newcommand{\qodd}{q^\mathrm{odd}}
\newcommand{\qeven}{q^\mathrm{even}}

The function $q_m(z)$ introduced in equation \eqref{eq:sh-baseline} is known
as the {\em(Associated) Legendre transform of order $m$},
\begin{equation}
  \label{eq:legendre-transform}
  \begin{split}
q_m(z_k) = \sum_{\ell = m}^\lmax \P^{m}_\ell(z_k) a_\lm,
    \end{split}
\end{equation}
assuming $m \ge 0$.  The following symmetry cuts the arithmetic
operations required in a SHT in half, as long as the spherical grid
distributes the rings symmetrically around the equator.  For any
non-negative integer $n$, the functions $\P_{m + 2n}^m(z)$ are even
and $\P_{m + 2n + 1}^m(z)$ are odd. We define $\qeven_m$ and $\qodd_m$
so that $\qeven_m$ contains the even-numbered and $\qodd_m$ the
odd-numbered terms of equation \eqref{eq:legendre-transform}, and so
that
\begin{equation}
  \label{eq:q-even-odd}
  q_m(z) = \qeven_m(z) + \qodd_m(z).
\end{equation}
Then, since $\qeven_m$ and $\qodd_m$ are weighted sums of even
and odd functions, respectively, they are themselves even and odd,
so that $q_m(-z)$ can be computed at the same time essentially for free,
\begin{equation}
  \label{eq:q-even-odd2}
  q_m(-z) = \qeven_m(z) - \qodd_m(z).
\end{equation}

For spherical harmonic analysis, one uses the orthogonality
property. Assuming $m \ge 0$,
\begin{equation}
\int \P_\ell^{m}(z) \P_{\ell'}^{m}(z) dz = \delta_{\ell \ell'},  
\end{equation}
so that
\begin{equation}
  \label{eq:backward-legendre}
  a_\lm = \int \P_\ell^{m}(z) q_m(z) dz.
\end{equation}
As discussed in Section \ref{sec:analysis}, the resulting quadrature
used in calculations can be exact or approximate, depending
on the placement of the pixel rings.
One can also in this case cut computation time in half by treating
even and odd $\ell - m$ separately.

\section{Fast Legendre Transforms}
\label{sec:fast-legendre}
As the Fourier transform part is essentially a solved problem, efforts
to accelerate SHTs revolve around speeding up the Legendre
transforms.  Let us write equation \eqref{eq:legendre-transform} as
\begin{equation}
  \label{eq:legendre-transform-linalg}
  \ve{q} = \L^T \ve{a},
\end{equation}
where we leave $m$ and the odd versus even case implicit. For a full
SHT, such a product must be computed for each of $2 (\lmax + 1)$ different $\L$
matrices. The backwards Legendre transform required for spherical harmonic
analysis is similarly
\begin{equation}
  \label{eq:legendre-transform-linalg-backwards}
  \ve{a} = \L \ve{q},
\end{equation}
give or take a set of quadrature weights.

The idea of Fast Legendre Transform algorithms is to compute equations
\eqref{eq:legendre-transform-linalg} and
\eqref{eq:legendre-transform-linalg-backwards} faster than $O(\lmax
N_\mathrm{ring})$. The approach of \cite{tygert3} is to factor $\L$ as
a product of block-diagonal matrices in a precomputation step, which
can significantly reduce the number of elements in total. This
technique is known as {\em butterfly compression}, and was introduced
by \cite{butterfly}.  The accuracy of the
compression is tunable, but even nearly loss-less compression with
close to double precision accuracy is able to yield significant gains
as the resolution increases.  We review the algorithm below, but
stress again that the reader should consult \cite{tygert3} for the
full details.  The butterfly compression technique was introduced by,

\subsection{The interpolative decomposition}
\newcommand{\IPA}{\widetilde{\A}} The core building block of the
compression algorithm is the Interpolative Decomposition (ID),
described in \cite{doi:10.1137/030602678}. Assume that an $m \times n$
matrix $\A$ has rank $k$, then the ID is
\begin{equation}
  \label{eq:ID}
  \A = \A^{(k)} \widetilde{\A},
\end{equation}
The matrix $\A^{(k)}$, known as the {\em skeleton matrix}, consists of
$k$ columns of $\A$, whereas $\widetilde{\A}$, the {\em
  interpolation matrix}, interpolates the eliminated columns from the
ones that are preserved.  Of course, $k$ of the columns of
$\widetilde{\A}$ must form the identity matrix.

The ID is obviously not unique; the trick is to find a decomposition
that is numerically stable. The algorithm of
\cite{doi:10.1137/030602678} finds an interpolation matrix
$\IPA$ so that no element has absolute value greater
than $2$, all singular values are larger than or equal to $1$, and the
spectral norm is bounded by $\sqrt{4 k (n - k) + 1}$.  The numerical
precision of the decomposition is tunable, as the decomposition found
by the algorithm satisfies
\begin{equation}
  \label{eq:ID-approximation}
  \| \A - \A^{(k)} \widetilde{\A} \|
 \le \sqrt{4 k (n - k) + 1} \sigma_{k + 1}
\end{equation}
where $\sigma_{k + 1}$ is the $(k+1)$ greatest singular value of $\A$.
Implementing lossy compression is simply a matter of reducing the
accuracy required of the IDs we use.

\newcommand{\T}{\ve{T}}
\newcommand{\B}{\ve{B}}
\newcommand{\Tk}[1]{\ve{T}^{(k)}_{#1}}
\newcommand{\Bk}[1]{\ve{B}^{(k)}_{#1}}
\newcommand{\TP}[1]{\widetilde{\ve{T}}_{#1}}
\newcommand{\BP}[1]{\widetilde{\ve{B}}_{#1}}

\subsection{Butterfly matrix compression}
\label{sec:butterfly}
\begin{figure*}
\centering
\includegraphics{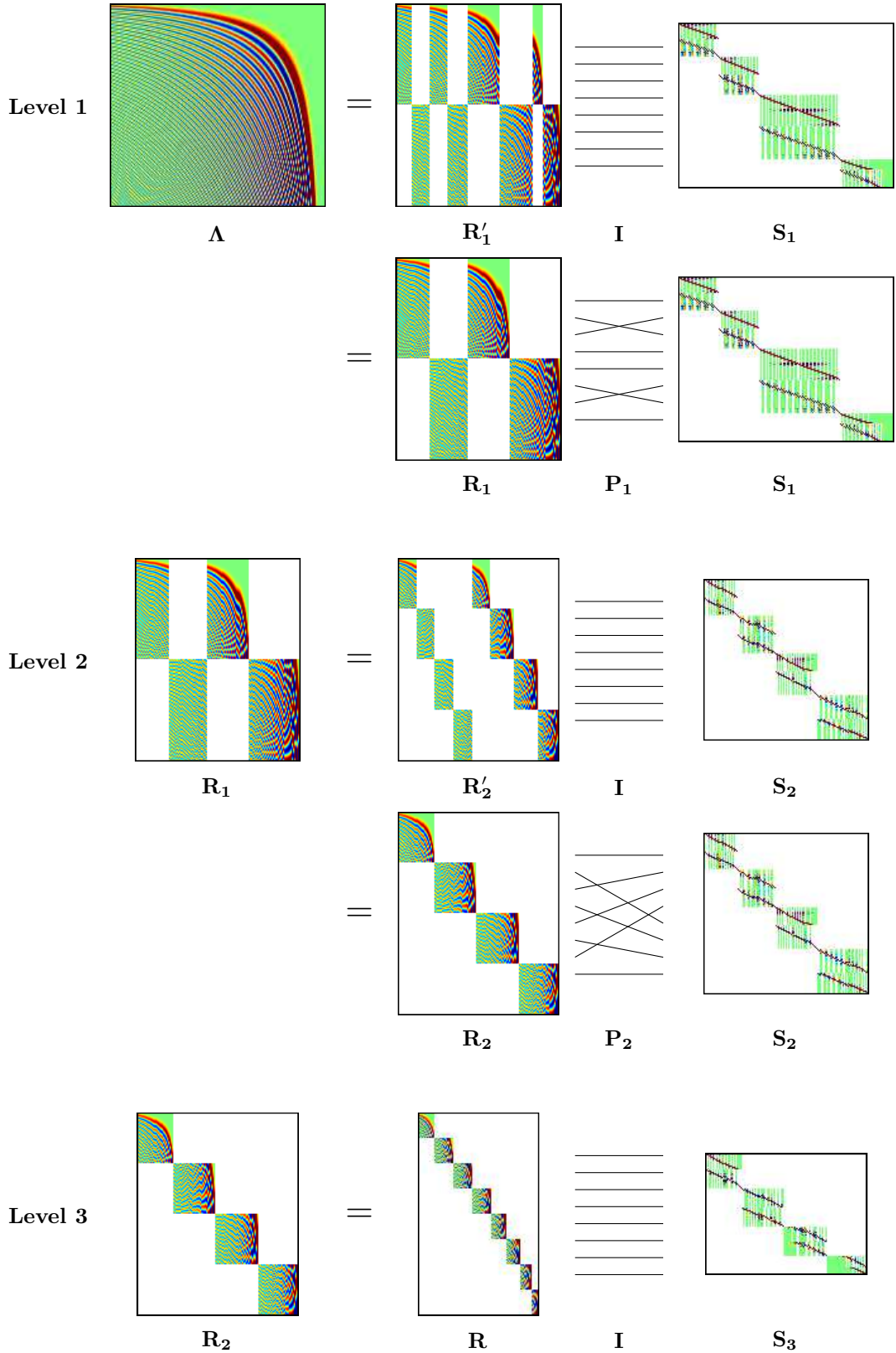}
\caption{Illustration of the butterfly matrix compression scheme.  On
  the first level, we use the Interpolative Decomposition to compress
  sub-blocks of the matrix $\L$ and produce the factorization $\L =
  \ve{R}'_1 \ve{S}_1$, where all blocks in $\ve{R}'_1$ have full
  rank. We then proceed by permuting the columns
  of $\ve{R}'_1$ so that $\L = \ve{R}_1 \ve{P}_1 \ve{S}_1$, in order to create new
  rank-deficient blocks.  The contents of the $\ve{S}_1$ matrix is
  saved as precomputed data, while we carry $\ve{R}_1$ along for
  further compression on the next level. The algorithm continues
  in this fashion until the residual matrix $\ve{R}$ only consists of a single
  diagonal of full-rank blocks. The final factorization becomes $\L =
  \ve{R} \ve{S}_3 \ve{P}_2 \ve{S}_2 \ve{P}_1 \ve{S}_1$.  The
  permutations involved are known in the FFT literature as {\em
    butterfly permutations}; the ``butterfly'' can be seen twice in
  the pattern of $\ve{P}_1$.}
\label{fig:butterfly}
\end{figure*}

We now use the ID recursively to factor the matrix $\L$.
After applying $p$ levels of compression, we have
\begin{equation}
\label{eq:factor-into-block-diagonal}
\L = \ve{R} \ve{S}_p \ve{P}_{p-1} \ve{S}_{p-1} \cdots \ve{P}_2 \ve{S}_2 \ve{P}_1 \ve{S}_1,
\end{equation}
where $\ve{R}$ is a block-diagonal {\em residual matrix} containing
elements that were not compressed, the $\ve{S}_i$ are block-diagonal
matrices containing compressed data, and the $\ve{P}_i$ are
permutation matrices. See Figure \ref{fig:butterfly} for an
illustration.  The structure of the permutations are very similar to
the {\em butterflies} used in FFT algorithms, hence the name of the
compression scheme. In fact, if one lets $\ve{S}_i$ contain a specific
set of $2\times2$-blocks on their diagonals one recovers the famous
Cooley-Tukey FFT.  In our case the blocks will be significantly
larger, typically around $150\times150$, although with much variation.

We start by partitioning $\L$ into $2^p$ column blocks. The number of
levels $p$ is mainly determined by the number of columns in the
matrix, so that the column blocks all are roughly of the same
predetermined width. In our case, $64$
columns worked well.

We then split each block roughly in half horizontally, and compress
each resulting block using the ID,
\begin{equation*}
  \label{eq:compression}
\begin{split}
  \L &= \mat{ \T_{1,1} & \T_{1,2} & \dots \\
              \B_{1,1} & \T_{1,2} & \dots } \\
&= \mat{ (\Tk{1,1} \cdot \TP{1,1}) & (\Tk{1,2} \cdot \TP{1,2}) & \dots \\
          (\Bk{1,1} \cdot \BP{1,1}) & (\Bk{1,2} \cdot \BP{1,2}) & \dots },
\end{split}
\end{equation*}
where the first subscript of each matrix refers to this being the first
iteration of the algorithm. It is useful to write the above matrix
as
\begin{equation*}
\begin{split}
\L
&= \mat{ \Tk{1,1}  & & \Tk{1,2}  & & \cdots \\
            & \Bk{1,1} & & \Bk{1,2} & \cdots}
   \mat{ \TP{1,1} &  &  \\
         \BP{1,1} &  &  \\
         & \TP{1,2} &   \\
         & \BP{1,2} &   \\ 
         & & \ddots }.
\end{split}
\end{equation*}
We denote the right matrix $\ve{S}_1$. It can not be further processed
and its blocks are simply saved as precomputed data, making use of the
fact that each block embeds the identity matrix in a subset of its columns.

The left matrix can be permuted and further compressed. For some
permutation matrix $\ve{P}_1$ we have:
\begin{equation*}
\begin{split}
\L
&= \mat{ \Tk{1,1}  & & \Tk{1,2}  & & \cdots \\
            & \Bk{1,1} & & \Bk{1,2} & \cdots}
   \ve{S}_1 \\
&= \mat{ \Tk{1,1}  & \Tk{1,2}  & &  & \cdots \\
            & & \Bk{1,1} & \Bk{1,2} & \cdots}
   \ve{P}_1 \ve{S}_1.
\end{split}
\end{equation*}
Then we join blocks horizontally, split them vertically, and compress
each resulting block. For the top-left corner we have
\begin{equation}
\label{eq:joinsplit}
\begin{split}
\left[ \begin{array}{cc} \Tk{1,1} & \Tk{1,2} \end{array} \right] =
\left[ \begin{array}{cc} \ve{T}_{2,1} \\ \ve{B}_{2,1} \end{array} \right] = 
\left[ \begin{array}{cc} (\Tk{2,1} \cdot \TP{2,1}) \\
                          (\Bk{2,1} \cdot \BP{2,1}) \end{array} \right].
\end{split}
\end{equation}
Applying this to all blocks in the matrix, we get
\begin{equation*}
\begin{split}
\L =& \mat{ 
  (\Tk{2,1} \cdot \TP{2,1})  &  &  & \cdots \\
  (\Bk{2,1} \cdot \BP{2,1})  &  &  & \cdots \\
   & & (\Tk{2,2} \cdot \TP{2,2}) & \cdots \\
   & & (\Bk{2,2} \cdot \BP{2,2}) & \cdots}
   \ve{P}_1 \ve{S}_1 \\
=&
\mat{ 
  \Tk{2,1}  &  &  & & \Tk{2,3} & & & & \cdots \\
   & \Bk{2,1}   &  & &   & \Bk{2,3} & & & \cdots \\
   & & \Tk{2,2}  & & & & \ddots & & \cdots \\
   & & & \Bk{2,2}  & & & &  & \cdots} \\ & \cdot
   \mat{ \TP{2,1} &  &  \\
         \BP{2,1} &  &  \\
         & \TP{2,2} &   \\
         & \BP{2,2} &   \\ 
         & & \ddots }
   \ve{P}_1 \ve{S}_1.
\end{split}
\end{equation*}
And so the scheme continues. For each iteration the number of
diagonals in the left matrix is halved, the number of blocks in
each diagonal is doubled, and the height of each block is roughly
halved. Eventually the left matrix consists only of a single diagonal
band of blocks, and further compression is impossible. This becomes
the residual matrix $\ve{R}$ of equation \eqref{eq:factor-into-block-diagonal}.

The efficiency of the scheme relies on the non-trivial requirement
that the $\Tk{}$ and $\Bk{}$ blocks are rank-deficient at every level
of the algorithm. To get a handle on which matrices exhibit this
behavior, we start with assuming the {\em rank property}, namely that
any contiguous rectangular sub-block of $\L$, up to the numerical
precision chosen, has rank proportional to the number of elements in
the sub-block. That is, the rank does not depend on the location or
shape of the block. Now, each time the butterfly algorithm joins two
skeletons, such as $[\Tk{1,1} \; \Tk{1,2}]$ in equation
\eqref{eq:joinsplit}, the resulting matrix has roughly $2k$ columns
while spanning out a corresponding block of $\L$ of rank
$k$. Therefore, half of the columns can be eliminated by applying the
ID. Since the data volume is roughly halved at each compression level,
and since $\ve{S}_i$ at each level has $O(\lmax)$ interpolative
matrices of size roughly $k \times 2k = O(1)$, the resulting
compressed representation of $\L$ has $O(\lmax \log \lmax)$
elements. See \cite{tygert3} for a more detailed argument.

\begin{figure*}
\centering
\includegraphics{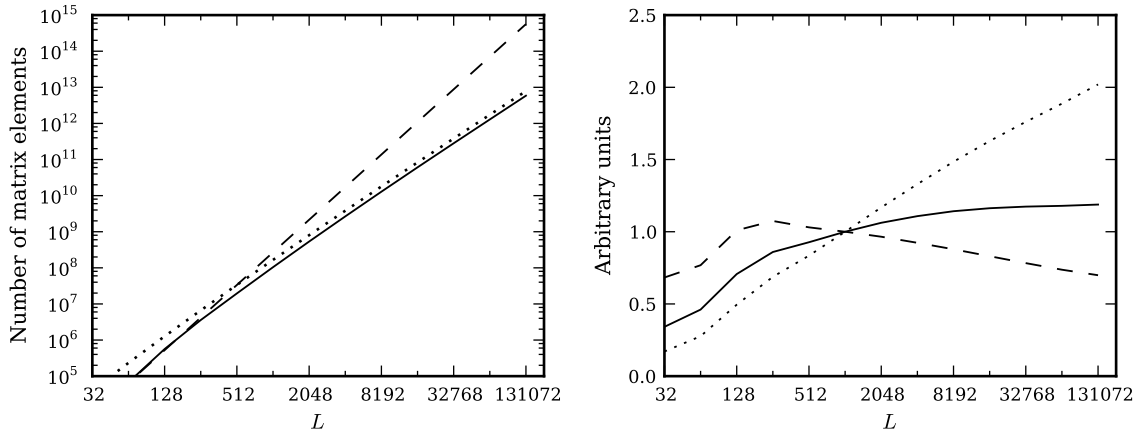}
\caption{Efficiency of the butterfly compression scheme.
  {\em Left panel:} Estimated size of compressed data (solid)
  compared to uncompressed matrix size (dashed). A line
  proportional to $O(\lmax^2 \log^2\lmax)$ (dotted) is shown for comparison.
  In each case we use a HEALPix grid with $\Nside = \lmax/2$.
  {\em Right panel:} A closer look on the computational
  scaling. The size of the compressed data is shown divided by
  $O(\lmax^2 \log^3 \lmax)$ (dashed), $O(\lmax^2 \log^2 \lmax )$
  (solid) and $O(\lmax^2 \log \lmax)$ (dotted), using an arbitrary
  normalization.}
\label{fig:prospects}
\end{figure*}

\cite{doi:10.1016/j.acha.2009.08.005} proves the rank property in the
case of Fourier transforms and Fourier-Bessel transforms. It is
however not proven in the case of associated Legendre functions
$\P_\ell^m(z)$. Figure \ref{fig:prospects} shows our results
for resolutions up to $\lmax \sim 130000$; we discuss these results
further in Section \ref{sec:predictions}.

\subsection{Notes on interpolation}
\label{sec:interpolation}
\cite{tygert2} describes an elegant and exact interpolation scheme which, in the
case of the HEALPix grid and $\lmax = 2 \Nside$, reduces the number of
required evaluation points for $q_m(z_k)$ by $2/3$.  Although our
conclusion was not to include this step in our code, we include a
brief discussion in order to motivate our decision.

We focus on the even Legendre functions, the odd case is similar.  Let
$n$ be an integer such that $\lmax < m + 2n$. The function
$P_{m+2n}^m(x)$ has $n$ roots in the interval $(0, 1)$, which we
denote $z_1, \dots, z_n$.  Now, assuming that we have evaluated $\qeven_m$
in these roots, we can interpolate to any other point $y \in (-1, 1)$
by using the formula\begin{equation}
  \label{eq:interpolation}
  \qeven_m(y) = \omega(y) \sum_{i=1}^{n}
    \frac{ \gamma(z_i) }{y^2 - z_i^2} \qeven_m(z_i),
\end{equation}
for some precomputed weights $\omega(y)$ and $\gamma(z_i)$.  The proof
relies on the Christoffel-Darboux identity for the normalized
associated Legendre functions
\citep{tygert2,doi:10.1006/jcph.1997.5782}.  The Fast Multipole Method
(FMM) allows the computation of equation \eqref{eq:interpolation} for
$p$ points with operation count of order $O(p + n)$ rather than 
$O(pn)$. The FMM was originally developed for accelerating $N$-body
simulations, but is here motivated algebraically. For more information
about one-dimensional FMM we refer to
\cite{doi:10.1137/S0036142997329232} and \cite{doi:10.1137/0733082}.

The reason we did not include this step in our code is that much of
the interpolation is already embedded in the butterfly matrix
compression.  Consider for instance $\Nside=2048$, $\lmax=3\Nside$ and
$m=2000$.  The full matrix $\L$ occupies 65 MiB when evaluated in the
HEALPix co-latitude nodes, and only 49 MiB when evaluated in the
optimal nodes as described above.  However, after compression the
difference is only 10.4 MiB versus 9.4 MiB. Thus the butterfly
compression compensates, at least partially, for the over-sampling.
Indeed, \cite{doi:10.1137/060662253} use a strongly related matrix
compression technique to implement the FMM itself.

Interpolation also causes
the precomputed data to become independent of the chosen
grid and resolution. However, we found the constant prefactor in the
FMM to be quite high, and including it only as a matter of convenience
appears to be out of the question for our target resolutions.  Since
the FMM has a linear computational scaling, the question
should be revisited for higher resolutions.

\subsection{CPU and memory trade-offs}
\label{sec:hybrid}

So far we have focused on reducing the number of floating point
operations (FLOPs).  However, during the past decade the speed of the CPU has
increased much more rapidly than the system memory bandwidth, so that
in current multi-core computers it is easy to get in a situation where
the CPUs are starved for data to process. When processing only one or
a few transforms concurrently, the volume of the precomputed data is
much larger than the volume of the maps being transformed, so that the
limitation is moving the precomputed data over the memory bus, not
processing power. Note that in the case of very many simultaneous
transforms the problem is alleviated since the movement of precomputed
data is amortized. Following in the footsteps of libpsht, and our own
requirements in CMB analysis, we have restricted our attention to
between one and ten concurrent transforms. While the butterfly
algorithm probably performs well in the face of many concurrent
transforms, it would require additional blocking and optimization
beyond what we have implemented, so that movement of the working
set in memory is properly amortized.  Note that as each $m$ is processed
independently, the working set is only about $1/\lmax$ of the total
input.

\begin{figure}[t]
\includegraphics{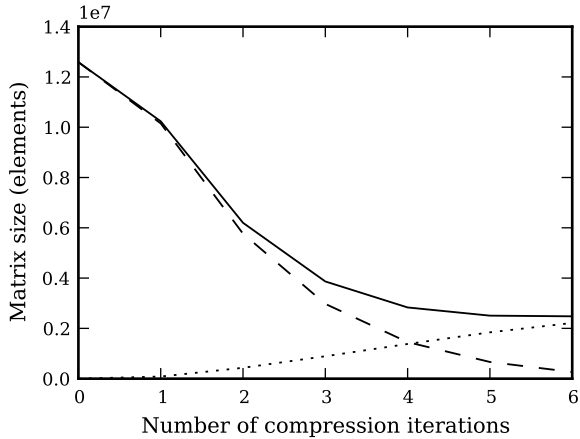}
\caption{Effect of each level of butterfly compression. The size of
  the compressed data (solid) is the sum of elements in residual
  uncompressed blocks in $\ve{R}$ (dashed) and the interpolation
  matrices $\ve{S}_i$ (dotted).  While $\ve{R}$ can be generated on
  the fly during transforms, the $\ve{S}_i$ needs to be stored as
  precomputed data, so that the choice of compression level is a
  trade-off between CPU use and the size of the precomputed data.
  Parameters for this figure are $\Nside=2048$, $\lmax=3\Nside$,
  $m=0$, and the initial chunk size $32$ columns.}
\label{fig:compression-efficiency}
\end{figure}

The considerations above motivates stopping compression early, after a
significant reduction in the floating-point operation  count has
been achieved, but before the size of the precomputed data becomes too
large (see Figure \ref{fig:compression-efficiency}).
Butterfly compression has the convenient feature that the
blocks in the residual matrix $\ve{R}$ consists of contiguous slices
from columns of $\L$. By orienting $\L$ so that rows are indexed by
$\ell$ and columns by $z$, the elements of the residual blocks can be
computed on the fly from three-term recurrence formulas for the
associated Legendre functions. We return to this topic in
Section \ref{sec:optimize-legendre}.

As an example, consider $\Nside=2048$ and $m=400$. The uncompressed
matrix $\L$ takes 64 MB in double precision. This can be compressed to
20\% of the original size by using five levels of compression, with
the uncompressed residual $\ve{R}$ accounting for about 13\% of the
compressed data. If one instead stops after three levels of
compression, then although the size of the compressed data has now
grown to 24\% of the original, 57\% of this is made out of elements in
$\ve{R}$. Since one only needs to store two elements for every column
of 512 elements in $\ve{R}$ and can generate the rest on the fly,
stopping compression after three levels reduces the memory bus traffic
and size of precomputed data by about 40\%, at the cost of some extra
CPU instructions. Note that the brute-force codes may simply be seen
as the limit of zero levels of compression.

\section{Implementation \& results}
\label{sec:implementation}

\subsection{Technology}

The Wavemoth library is organized in a core part and an auxiliary
part. The core is primarily written in C and contains the routines for
performing spherical harmonic transforms. The auxiliary shell around
the core is written as a Python package, and is responsible generating
the precomputed data using the butterfly compression algorithm, as
well as the regression and unit tests.

By writing the core in pure C we remain close to the hardware, and
make sure the library can be used without Python.  C remains the
easiest language to call from other languages such as Fortran, C++,
Java, Python, MATLAB, and so on. By using Python in the auxiliary
support code we accelerate development of the parts that are not
performance critical, and make writing tests a pleasant experience.
Being able to quickly write up unit tests is an indispensable tool, as
it allows optimizing the C code iteratively without introducing
bugs. Since individual pieces of the C core is tested, there is both a
public API for end-users and a private API that is used from Python to
test individual C routines in isolation.  Much of the support code is
implemented in Cython \citep{doi:my/cython}, which bridges the worlds
of Python and C.

The C core depends on files containing precomputed data, a Fourier
transform library and a BLAS library. For the latter two we use FFTW3
\citep{doi:my/fftw3} and ATLAS \citep{doi:my/atlas}, respectively.
Parts of the Wavemoth core is written using templates in order to
generate many slight variations of the same C routine. We use
Tempita\footnote{%
  http://pythonpaste.org/tempita/}, a purely text-oriented templating
language, and find this to be much more convenient for optimizing a
computational core than the type-oriented templates of C++.
During the precomputations, we use the open source Fortran 77 library
ID\footnote{http://cims.nyu.edu/\~{}tygert/software.html} to
compute the Interpolative Decomposition, and libpsht to generate the
associated Legendre functions.

Unlike libpsht, we have not focused on portability, and Wavemoth is
only tested on 64-bit Linux with the GCC compiler on Intel-platform
CPUs.  Computational cores are written using SSE intrinsics and
128-bit registers. More work is needed for optimal performance on the
latest Intel micro-architecture, which support 256-bit registers, or
on non-Intel platforms.  Beyond that, we expect no hurdles in
improving portability.

\subsection{Benchmarks}
\label{sec:benchmarks}

\begin{figure*}
\epsscale{1}
\plotone{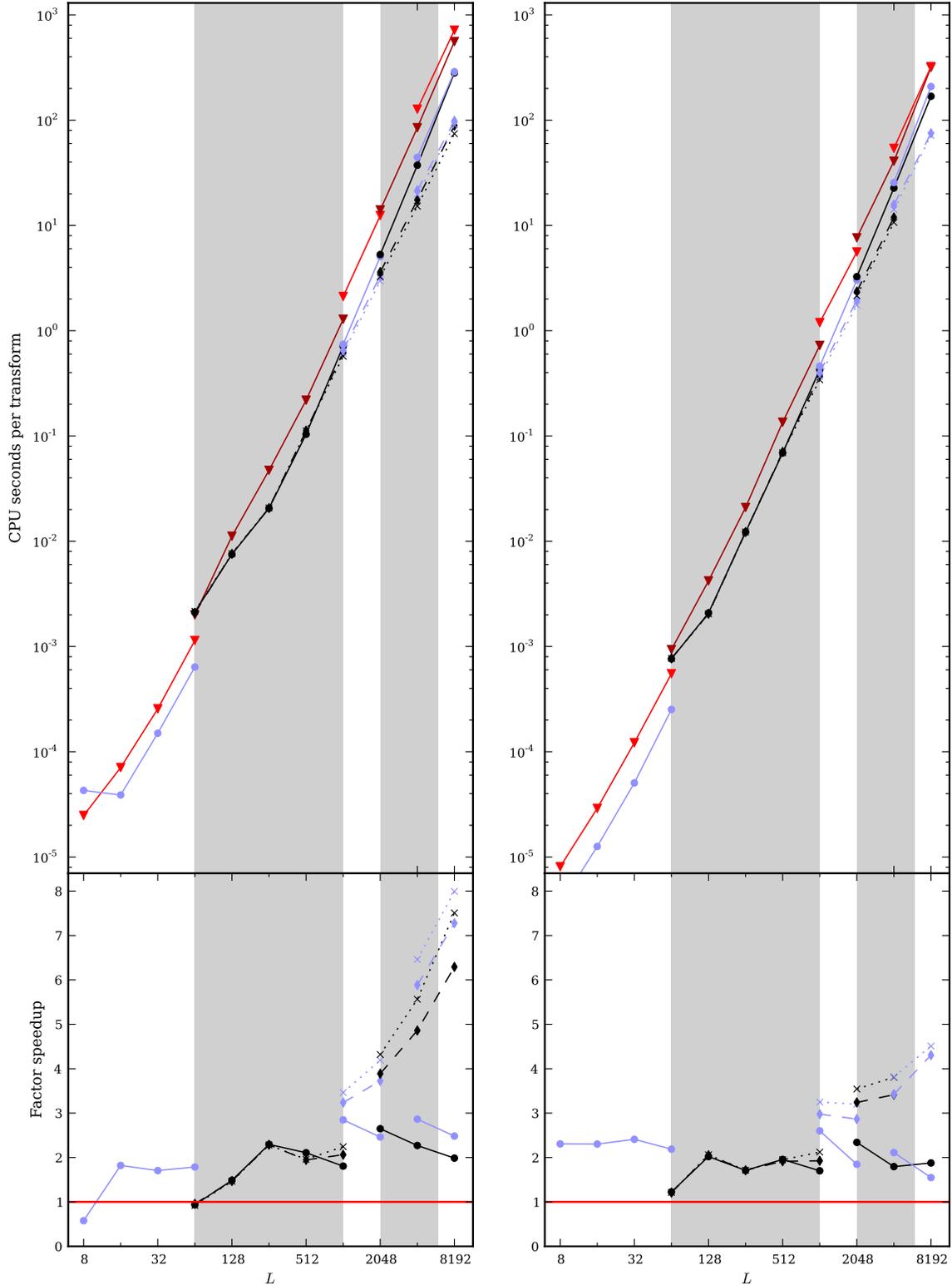}

\caption{Benchmarks for full SHTs performed on the Intel system. The left pane shows
  timings for a single transform, the right for ten simultaneous
  transforms. We scale up the number of CPU cores together with the
  resolution. Each pane is divided into four partially overlapping
  segments corresponding to 1, 8, 16, 32, and 64 CPU cores,
  respectively (indicated by white/gray backgrounds and changes in line
  colors).  The libpsht code (red triangles) is compared to Wavemoth
  (blue/black) with no compression (solid, circles), compression with
  precision $10^{-13}$ (dashed, diamonds) and compression with precision
  $10^{-8}$ (dotted, crosses). In each case we use a HEALPix grid with
  resolution $\Nside=\lmax/2$.
  Note for instance how both codes suffer from parallelization overhead
  at the transition from one to eight cores, but that libpsht suffers
  less and catches up with Wavemoth. For a single transform at
  high resolutions, the situation is the contrary, with Wavemoth
  parallelizing better at the jump from 16 to 32 cores and
  from 32 to 64 cores.
  We repeated each benchmark multiple times both with and without
  HyperThreading, and report the fastest wall clock time achieved
  multiplied with the number of CPU cores used and divided by the
  number of simultaneous transforms. Some 32-core timings for ten
  simultaneous transforms at $\lmax=8192$ could not be obtained due to
  memory limitations. The load time of the precomputed data from the
  hard drive is not included.}
\label{fig:intel}
\end{figure*}

\begin{figure*}
\centering
\includegraphics{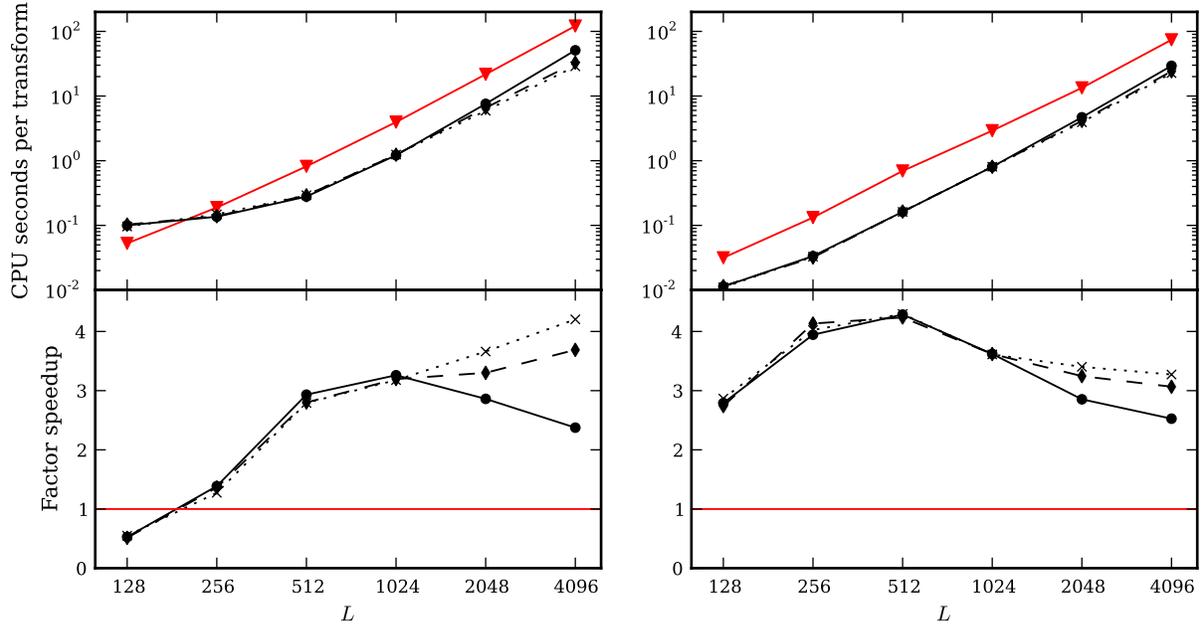}
\caption{Benchmarks for full SHTs performed on the AMD system, using
  all 48 CPU cores.  The libpsht code (red triangles) is compared to
  Wavemoth (blue) with no compression (solid, circles), compression
  with precision $10^{-13}$ (dashed, diamonds), compression with
  precision $10^{-8}$ (dotted, crosses).  Left pane shows a single
  transform and the right pane ten simultaneous transforms. In each
  case we use a HEALPix grid with resolution $\Nside=\lmax/2$.  The
  large speedup in the range $\lmax=256..1024$ is in part due to
  Wavemoth scaling better to all 48 cores, and is closer to a 2x
  speedup when using fewer cores.  We repeated each benchmark multiple
  times, and report the fastest wall clock time achieved multiplied
  with the number of CPU cores used and divided by the number of
  simultaneous transforms. }
\label{fig:amd}
\end{figure*}

\begin{deluxetable*}{rrrrrrc}
\tabletypesize{\small}
\tablecaption{\label{tab:sizes}Size of precomputed data}
\tablecomments{The precomputation time quoted is
the wall time taken to compute at tolerance $10^{-13}$ on the Intel
system, multiplied by the number of CPU cores used.
We use 1 core for $L=32$, and then scale up gradually to 64
cores at $L=8192$. The precomputed data is saved to a network file system.}
\tablecolumns{5}
\tablehead{
  \colhead{$\lmax$} &
  \colhead{No comp.} &
  \multicolumn{2}{c}{Intel ($\rho=7.5$)} &
  \multicolumn{2}{c}{AMD ($\rho=18$)} &
  \colhead{Precomputation time}
  \\
  \colhead{} &
  \colhead{} &
  \colhead{Tol. $10^{-13}$} &
  \colhead{Tol. $10^{-8}$} &
  \colhead{Tol. $10^{-13}$} &
  \colhead{Tol. $10^{-8}$} &
  \colhead{(CPU minutes)}}
\startdata
32   & 130 KiB & --     & --       &  --     & --      & 0.02 \\
64   & 496 KiB & --     & --       &  --     & --      & 0.03 \\
128  & 2.0 MiB & 2.0 MiB& 2.0 MiB  & 1.9 MiB & 1.9 MiB & 0.22 \\
256  & 8.0 MiB & 8.0 MiB& 8.0 MiB  & 7.1 MiB & 7.1 MiB & 2.6 \\
512  & 27 MiB  & 174 MiB& 187 MiB  & 27 MiB  & 27 MiB  & 7.4 \\
1024 & 102 MiB & 937 MiB& 988 MiB  & 102 MiB & 170 MiB & 12 \\
2048 & 389 MiB & 6.0 GiB& 5.8 GiB  & 4.4 GiB & 4.3 GiB & 90 \\
4096 & 1.5 GiB & 38 GiB & 35 GiB   & 28 GiB  & 27 GiB  & 536 \\
8192 & 5.8 GiB & 212 GiB & 208 GiB & --      &  --     & 4380
\enddata
\end{deluxetable*}

\begin{deluxetable*}{rccc}
\tabletypesize{\small}
\tablecaption{\label{tab:accuracy}Samples of numerical accuracy}
\tablecomments{In each case, the transform of a single Gaussian sample
is compared against libpsht double precision results.}
\tablecolumns{4}
\tablehead{\colhead{$\lmax$} & \colhead{No compression} & \colhead{Tolerance $10^{-13}$} &
  \colhead{Tolerance $10^{-8}$}}
\startdata
8 & 8.6e-16 & -- & -- \\
16 & 1.4e-15 & -- & -- \\
32 & 2.7e-15 & -- & -- \\
64 & 5.8e-15 & -- & -- \\
128 & 1.2e-14 & -- & -- \\
512 & 5.1e-14 & 4.7e-14 & 1.9e-09 \\
1024 & 1.3e-13 & 8.9e-14 & 2.4e-09 \\
2048 & 2.7e-13 & 1.7e-13 & 3.1e-09 \\
4096 & 6.4e-13 & 3.3e-13 & 3.7e-09 \\
8192 & 2.2e-12 & 6.6e-13 & 4.2e-09
\enddata
\end{deluxetable*}

We include benchmarks for two different systems with different memory
bandwidth, as Wavemoth's performance is deeply influenced by this
aspect of the hardware.  Figure \ref{fig:intel} presents benchmarks
taken on a 64-core 2.27 GHz Intel Xeon X7560 (Nehalem
micro-architecture), which has a compute-to-bandwidth ratio of about
45:1. Figure \ref{fig:amd} presents benchmarks taken on a 48-core 2.2
GHz AMD Opteron 6174. The compute-to-bandwidth ratio is in this case
about 64:1, significantly worse than the Intel system\footnote{The
  Intel system supports transfer of 13 billion numbers per second and
  has theoretical peak compute power 580 GFLOPS, using all 64
  cores. The AMD system supports transfer of 6.5 billion numbers per
  second and has theoretical peak compute power of 422 GFLOPS, using
  all 48 cores. All numbers refer to double precision floating
  point.}.  The consequence is that butterfly compression gives less
of an advantage, with only about four times speedup over libpsht at
$\lmax=4096$, compared to the corresponding six times speedup achieved
on the Intel system. In the case of ten simultaneous transforms,
libpsht achieves a very consistent 2x speedup which Wavemoth is not
able to fully match, as most of our tuning effort has been on the
single transform path.

The highest tested accuracy of $\epsilon=10^{-13}$ for the Legendre
transforms was chosen because current codes using the HEALPix grid
only agree to this accuracy on high resolutions
\citep{libpsht}.

An important aspect of the systems for our purposes is the non-uniform
memory access (NUMA). On each system, the CPU cores are grouped into
eight nodes, and the RAM chips evenly divided between the nodes.  Each
CPU only have direct access to RAM chips on the local node, and must
go through a CPU interconnect bus to access other RAM chips. For
consistent performance we need to ensure that Wavemoth distributes
the precomputed data in such a way that each CPU finds the data it
needs in its local RAM chips.  In the benchmarks we always use a whole
number of nodes, so that computation power and memory bandwidth scale
together. The exception is benchmarks using a single core, but in
those cases, Wavemoth's precomputed data fits in cache anyway.

Table \ref{tab:sizes} list the sizes of the precomputed data.  To
balance bandwidth and CPU requirements as described in Section
\ref{sec:hybrid}, the precomputation code takes a parameter $\rho$,
specifying the cost of floating point operations in the
bandwidth-intensive butterfly matrix application stage relative to the
cost of floating point operations in the CPU-intensive brute-force
Legendre transform stage. The parameter was then tuned for the
single-transform case for $\lmax=4096$, resulting in optimal choices
of $\rho=7.5$ on the Intel system and $\rho=18$ on the AMD system.
Performing the precomputations scales as $O(\lmax^3)$.  In the case of
no compression, we still store the precomputed quantities necessary
for the Legendre recurrence relations in memory, as described in the
appendix. Loading this data from memory is not necessarily faster than
computing it on the fly, but doing so saved some development time.

All methods involved are numerically stable and well understood, so we
do not include a rigorous analysis of numerical accuracy.  Table
\ref{tab:accuracy} lists the relative error from transforming a single
set of standard Gaussian coefficients per configuration.  We use the
relative error
\begin{equation}
\label{eq:relative-error}
\epsilon = \sqrt{\sum_{i=1}^\Npix (x_i - y_i)^2 / \sum_{i=1}^\Npix x_i^2},
\end{equation}
where $x_i$ denote the result of libpsht and $y_i$ the result of our
code.  The discrepancies in the no-compression, high-$\lmax$ cases are
due to using a different recurrence for the associated Legendre functions,
as described in Appendix \ref{sec:optimize-legendre}.  As we did not
compare with higher precision results, it is not clear whether it is
our code, libpsht, or both, that loose precision with higher
resolution. Note that the input data to the butterfly compression is
generated using libpsht.

\subsection{Higher resolutions}
\label{sec:predictions}

Due to memory constraints we have not gone to higher resolutions than
$\lmax \sim 8000$. Instead, we provide estimates for the number of
required floating point operations. \cite{tygert3} provide similar
estimates, but focus on the behavior for the Legendre transform for
single $m$ rather than the full SHT.

At each resolution, we compress $\L_m^\text{odd}$ for 20
different $m$, and fit the cost estimate
\begin{equation}
  \label{eq:costfit}
  \hat{c}_m = \alpha + \sum_{p=0}^2 \beta_p m \log(1+m)^p
\end{equation}
by least squares minimization in the parameters $\alpha, \beta_0,
\beta_1$ and $\beta_2$. The final cost is then estimated by
\begin{equation}
\label{eq:nflops}
\hat{c}_\mathrm{total} = 2 \sum_{m=0}^{\lmax} \hat{c}_m,
\end{equation}
since $\L_m^\text{even}$ and $\L_m^\text{odd}$ has almost
identical behavior. The results can be seen in Figure
\ref{fig:prospects}.  For $\lmax \sim 130000$, the butterfly algorithm
requires only 1\% of the arithmetic operations of a brute force
transform. The size of the precomputed data at this resolution is
around 45 TiB in double precision, although this can be reduced by
using the hybrid approach of Section \ref{sec:hybrid}.

At low resolutions, the algorithm is bound by the $O(\lmax^3)$
operations of the brute-force Legendre transform. At high resolutions,
the $O(\lmax^2 \log^2 \lmax)$ trajectory is clearly a better fit than
the $O(\lmax^2 \log \lmax)$ scaling conjectured by \cite{tygert3}. Note
that the numerical evidence presented in \cite{tygert3} show that the
average $k$ increases monotonically with $m$, so it may indeed be the
case that the rank property is not fully satisfied, or only satisfied
conditional on $m$.  The benchmark results of \cite{tygert3} seem
to be in agreement with the $O(\lmax^2 \log^2 \lmax)$ hypothesis as
well.

\begin{figure}[t]
\includegraphics{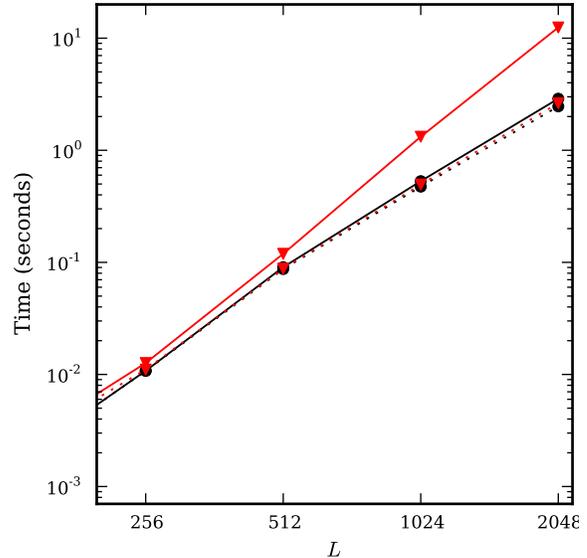}
\caption{Comparison of Wavemoth (black circles) with the algorithm of
  \cite{mohlenkamp} as implemented in libftsh (red triangles), with
  accuracy $10^{-13}$ (solid) and $10^{-8}$ (dotted). Both codes are
  run on a single core. Only the Legendre transform part is
  benchmarked, as libftsh does not implement the full SHT.
  Wavemoth uses a HEALPix grid with
  $2\lmax - 1$ rings, while libftsh uses a Gaussian grid with
  $2\lmax$ rings. The placement of the rings should make little
  difference to the performance of either code.}
\label{fig:mohlenkamp}
\end{figure}

\subsection{Comparison with other fast SHT algorithms}
\label{sec:comparison-with-others}
A widely known scheme for fast SHTs is the $O(\lmax^2 \log^2 \lmax)$
transform of \cite{doi:10.1007/s00041-003-0018-9}, implemented in
SpharmonicKit. It algebraically expresses a Legendre transform of
degree $\lmax$ as a function of two Legendre transforms of degree
$\lmax / 2$, resulting in a divide-and-conquer scheme similar to the
FFT algorithms. Unfortunately, the scheme is inherently numerically
unstable, and special stabilization steps must be incorporated. Also,
it is restricted to equiangular grids, so that it can not be used
directly with the HEALPix or GLESP grids.  \cite{doi:10.1086/507692}
benchmarks SpharmonicKit against the original HEALPix implementation
(pre 2.20) and find that it is almost three times slower at
$\lmax=1024$. Keep in mind that libpsht, used in present releases of
HEALPix, is about twice as fast as the original HEALPix
implementation.  Considering the above, we stop short of a direct
comparison between Wavemoth and SpharmonicKit. Note that while
SpharmonicKit achieves much higher accuracy of an SHT round-trip than
HEALPix does, this is an effect of the different sampling grids being
used, not of the computational method, and it is straightforward to
extend the Wavemoth code to use the same grid as SpharmonicKit.

\cite{mohlenkamp} uses a matrix compression technique similar to the
one employed in this paper, which is independent of the pixel grid
chosen.  A matrix related to the $\L$ of the present paper is locally
approximated by truncated trigonometric series. The resulting SHT
algorithm scales as $O(\lmax^{5/2} \log \lmax)$. As shown in Figure
\ref{fig:mohlenkamp}, the code behaves very similarly to our code at
medium resolution, as long as one do not require too much numerical
accuracy.  The size of the precomputed data is also of the same order,
sometimes half and sometimes double that of Wavemoth's data.

Note that libftsh appears to have potential for optimization for
modern platforms, and this should be taken into account when comparing
the algorithms. Due to its age, libftsh makes assumptions about 32-bit
array sizes which prevents comparison at higher resolutions without
porting libftsh to 64-bit. The libftsh code contains an implementation
of the Legendre transforms only, and not of the full spherical
harmonic transforms. It should be straightforward to modify Wavemoth
to use libftsh for its Legendre transforms in order to perform full
SHTs using this algorithm.

The compression scheme of \cite{mohlenkamp} appears to be very
competitive for low accuracy transforms, but less so if higher
precision is needed. It may be fruitful to hybridize the algorithms of
\cite{tygert3} and \cite{mohlenkamp} and use both together to compress
a single matrix. Even if that does not work, one can simply use
whichever performs best for a given $m$.

\section{Discussion}
There is significant potential in speeding up spherical harmonic
transforms beyond the codes in popular use today. We achieved a 2x
speedup at low and medium resolutions simply due to restructuring how
the brute-force computations are done, and believe there is potential
for even more speedup if time is spent on profiling and
micro-optimization. In particular, our code is under-optimized
for multiple simultaneous transforms.

At the highest resolutions in practical use in cosmology today, $L
\sim 4000$, use of the butterfly compression is borderline. One the
one hand, it does yield an additional 2x speedup; potentially much
more if one needs less accuracy. On the other hand, it requires
between 30 and 40 GiB of precomputed data in memory, and the
transportation of that data over the memory bus for every set of
transforms. The result is a delicate balance between bandwidth and
achieved speedup; for every number stored in the precomputed data, one
might save 40 arithmetic operations, but then again computation is
much cheaper than accessing system memory on present-day computer
architectures.

In Section \ref{sec:interpolation}, we note the existence of
interpolation schemes that cut the necessary sample points for
brute-force codes by two thirds in the case of the HEALPix grid;
although performing the interpolation step does not come for free. It
seems that the speedup from such interpolation alone could on the same
order as what the butterfly algorithm achieves for the current needs
of CMB research. The advantage is that it does not require
nearly as much precomputed data, and is so much less
architecture-dependent and easier to micro-optimize. In going forward
we therefore anticipate spending more effort on direct interpolation
schemes and less effort on matrix compression. For resolutions
higher than those needed in CMB analysis, matrix compression
schemes seem like the most mature option at the moment.

We have not discussed spin-weighted spherical harmonic transforms,
which are crucial to analyzing the polarization properties of the CMB.
However, \cite{doi:10.1006/jcph.2000.6551} and
\cite{doi:10.1016/j.jcp.2007.07.005} describe how the transform of a
polarized CMB map can be reduced to three scalar transforms.
This would additionally help amortize the memory bus transfer of the
precomputed data. Alternatively, it may be possible to compress the
spin-weighted spherical harmonic operators.

We consider Wavemoth an experimental code for the time being, and
spherical harmonic analysis has been left out. This was done purely to
save implementation time, and we know of no obstacles to implementing
this using the same methods. The code also lacks support for MPI
parallelization, although we expect adding such support to be
straightforward. The only inter-node communication requirement is a
global transpose of $q_m(z_k)$ between the Legendre transforms and the
Fourier transforms.

\begin{acknowledgements}
  The author thanks S. K. N\ae{}ss, H. K. Eriksen, M. Tygert, M. Reinecke
  and M. Mohlenkamp for useful discussions, and M. Omang and F. Hansen
  for lending the benchmark hardware. The author is funded by European
  Research Council grant StG2010-257080. The benchmark hardware is
  funded by the Norwegian Defence Estates Agency and the Research
  Council of Norway.
\end{acknowledgements}

\appendix

\section{Implementation details}

\subsection{Applying the compressed matrix representation to a vector}
On modern computers, the primary bottleneck is often to move data
around. Fundamental design decisions were made with this in mind.
Looking at the compressed representations of $\L$ in Section
\ref{sec:butterfly}, the immediate algorithm that comes to mind for
computing $\L\ve{x}$ or $\L^T\ve{x}$ is the breadth-first approach:
First compute $\ve{S}_1\ve{x}$, then permute the result, then compute
$\ve{S}_2(\ve{P}_1\ve{S}_1\ve{x})$, and so on. However, this leads to
storing several temporary results for longer than they need to, since
the rightmost permutations are very local permutations, and only the
leftmost permutation is fully global.  Therefore we traverse the data
dependency tree set up by the permutations in a depth-first
manner. The advantage of this approach is that it is {\em cache
  oblivious} when transforming a few vectors at the time. That is, it
automatically minimizes data movement for any cache hierarchy, whereas
breadth-first traversal will always drop to the memory layer that is
big enough to hold the entire set of input vectors.  Note that for
transforming many maps at the same time, cache-size dependent blocking
should be implemented in addition, but we have stopped short of
this. Like \cite{tygert3}, we also do the compression during
precomputation depth-first, which ensures that, per $m$, memory
requirements go as $O(\lmax \log \lmax)$ even though computation time
go as $O(\lmax^2)$.

The core computation during tree traversal is to apply the
interpolative matrices, e.g., $\TP{}\ve{x}$ or $\TP{}^T \ve{x}$. Keep
in mind that the $k$-by-$n$ matrix $\TP{}$ contains the $k$-by-$k$
identity matrix in a subset of its columns; making use of this is
important as it roughly halves the storage size and FLOP count. Given
an ID $\ve{T} = \Tk{} \TP{}$, we can freely permute the rows of
$\TP{}$, simply by permuting the columns of $\Tk{}$ correspondingly.
We do this during precomputation to avoid the unordered memory usage
pattern of arbitrary permutations. Instead, we can simply filter the
input or output vectors into the part that hits the identity
sub-matrix and the part that hits the dense sub-matrix.

\subsection{Efficient code for Legendre transforms}
\label{sec:optimize-legendre}

As mentioned in Section \ref{sec:hybrid}, it is necessary to balance
the amount of precomputed data to the memory bandwidth, so code is
required to apply the residual blocks in $\ve{R}$ to vectors without
actually storing $\ve{R}$ in memory.  This means computing
a cropped version of the Legendre transform,
\begin{equation}
  \label{eq:legendre-jump-two}
  \begin{split}
q'(z_j) = \sum_{k=k_\text{start}}^{k_\text{stop}} \P^{m}_{m + 2k + t}(z_j) a_{\ell m},
    \end{split}
\end{equation}
where $t = 0$ for the even transforms and $t = 1$ for the odd
transforms. To compute $\P_{\ell}^m$ we use a relation that jumps two
steps in $\ell$ for each iteration \citep{tygert3}:
\begin{equation}
\label{eq:three-term}
\begin{split}
\P_{\ell + 2}^m(z) &= \frac{z^2 - d^m_l}{c^m_\ell} \P_{\ell}^m(z) -
   \frac{c^m_{\ell - 2}}{c^m_\ell} \P_{\ell - 2}^m(z) \\
   &\equiv (z^2 + \alpha^m_\ell)\beta_\ell^m \P_{\ell}^m(z) + \gamma^m_\ell\P_{\ell - 2}^m(z),
\end{split}
\end{equation}
with
\begin{equation*}
  c^m_\ell = \sqrt{\frac{(\ell - m + 1)(\ell - m + 2)(\ell + m + 1)(\ell + m + 2)}
{(2\ell + 1)(2\ell +3)^2(2\ell + 5)}}
\end{equation*}
and
\begin{equation*}
  \label{eq:c_l}
  d^m_\ell = \frac{2\ell(\ell + 1) - 2m^2 - 1}{(2\ell - 1)(2\ell + 3)}.
\end{equation*}
This recurrence relation requires five arithmetic operations per
iteration, as opposed to a more widely used relation which takes one
step in $\ell$ and only needs four arithmetic operations per step
\citep[see, e.g.,][]{NR}. However, since $\L^\text{even}$ and
$\L^\text{odd}$ may have different columns in the residual blocks of
their compressed representations, relation \eqref{eq:three-term} is a
better choice in our case.

For each block in $\ve{R}$ we precompute $\alpha^m_\ell$,
$\beta^m_\ell$ and $\gamma_{\ell}^m$, as well
$\P_{k_{\mathrm{start}}}^m(z)$ and $\P_{k_{\mathrm{start}} + 1}^m(z)$
for each $z$ for initial conditions. Note that $\P_\ell^m(z)$ in parts
of its domain take values so close to zero that they can not be
represented in IEEE double precision.  However, in these cases
$\P_\ell^m(z)$ is always increasing in the direction of increasing
$\ell$, so we can simply increase $k_\text{start}$ correspondingly. In
fact, we follow libpsht and assume that the dynamic range of the input
data is small enough, within each $m$, that values of $\P_\ell^m(z)$
smaller than $10^{-30}$ in magnitude can safely be neglected.
As far as possible we group together six and six columns
with the same $k_\mathrm{start}$ and $k_\mathrm{stop}$, for reasons that will
soon become clear.

For an efficient implementation, the first important point is to make
sure the number of loads from cache into CPU registers is balanced
with the number of floating-point operations.  The second is to make
sure there are enough independent floating-point operations in flight
simultaneously, so that operations can be pipelined. Thus,
\begin{itemize}
\item for performing a single transform with one real and one
  imaginary vector, the values of $\P_\ell^m$ should never need to
  leave the CPU registers. Rather, we fuse equation
  \eqref{eq:legendre-jump-two} and equation \eqref{eq:three-term} in
  the core loop. For multiple simultaneous transforms we save
  $\P_\ell^m$ to cache, but make sure to process in small batches that
  easily fit in L1 cache.
\item we process for several $z_j$ simultaneously. This amortizes the
  register loads of $\alpha^m_\ell$, $\beta^m_\ell$ and
  $\gamma_{\ell}^m$. It also ensures that there are multiple
  independent chains of computation going on so that pipelining works
  well.
\end{itemize}
In the single transform case with one real and one imaginary vector,
we do the full summation for six $z_j$ at the time (when possible).
The allocation of the 16 available 128-bit registers, each holding two
double-precision numbers, then becomes three registers for
$\P_\ell^m$, three for $\P_{\ell-2}^m$, three for the auxiliary data
$\alpha^m_\ell$, $\beta^m_\ell$ and $\gamma_{\ell}^m$, six
accumulation registers for $q'(z_j)$, and one work register. The
$z_j^2$ values are, perhaps counter-intuitively, read again from cache
in each iteration, which conserves three registers and thus enables
processing six $z_j$ in each chunk instead of only four without
register spills. Finally, when the time comes for multiplying
$\P_\ell^m$ with $a_{\ell m}$, the auxiliary data is no longer needed,
leaving room for loading $a_{\ell m}$.

On the Intel Xeon system, the routine performs at
6.46 GFLOP/s per core (71\% of the theoretical maximum) when
benchmarked on all the Legendre transforms necessary for a full SHT
across 32 cores. The effect of instruction pipelining is evident;
reducing the number of columns processed in each iteration from six to
four reduces performance to 5.69 GFLOP/s (63\%), and when only
processing two columns at the time, performance is only
4.28 GFLOP/s (47\%).

We skip the details for the multiple transform case, but in short, in
involves the same sort of blocking performed for matrix
multiplication, including repacking the input data in
blocks. \cite{goto} provide an excellent introduction to
blocking techniques.  In this case the performance is 5.60 GFLOP/s
(62\%) per core when performing the Legendre transforms necessary for
ten simultaneous SHTs.


The considerations above guided the choice of loop structure, which
was then implemented in pure C using SSE intrinsics. We did not spend
much time on optimization, so there should be room for further
improvements, in particular for the multiple-transform path.

\subsection{Data layout}
The butterfly compression algorithm naturally leads to the following
code organization for spherical harmonic synthesis:
\begin{enumerate}
\item Since each $m$ is processed independently, we request input in
  $m$-major ordering. Also, for multiple simultaneous transforms, the
  coefficients of each map are interleaved, which is optimal both for
  the butterfly algorithm and the brute-force cropped Legendre
  transforms. In most places, the real and complex parts of the input
  can be treated as two independent vectors, since $\L$ is a real
  matrix.
\item Compute all $q_m(z_j)$ into a 2D array. Since each $m$ is
  processed independently, this ends up in $m$-major ordering,
  like the input.
\item While transposing the $q_m(z_j)$ array into ring-major ordering,
  phase-shift and wrap around the coefficients, and perform FFTs on
  each ring. Rings must be processed in small batches in order to
  avoid loading cache lines multiple times.
\end{enumerate}
A temporary work buffer with size of the same order as the input and
output is used for $q_m(z_j)$. An in-place code should be feasible
with the use of an in-place transpose.

A drawback compared to brute-force codes is that $q_m(z_j)$ needs to
first be written to and then read from main memory. Here, libpsht is
instead able to employ blocking, so that a few rings at the time are
completely processed before moving on. Our benchmarks do however
indicate that this is not a big problem in practice. Also, for cluster
parallelization using MPI, it would be natural to
follow S$^2$HAT \citep{s2hat-a,s2hat-b} in distributing the input
data by $m$ and the output data by rings, which also leads to a global
transpose operation.

Wavemoth stores the output maps in interleaved order, since FFTW3 is able to
deal well with such transforms. The libpsht code is able to
support any output ordering, although stacked, non-interleaved maps
are slightly faster, so that is the ordering we use for libpsht
in the benchmarks.

\end{document}